\documentclass[conference]{IEEEtran}
\usepackage{amssymb,amsmath,amsthm}
\usepackage{caption}
\usepackage{mdwmath}
\usepackage{mdwtab}
\usepackage{soul}
\usepackage{cite}
\usepackage{multirow}
\usepackage{color} 
\usepackage{setspace}
\usepackage{esvect}
\usepackage{amsmath}
\usepackage{algorithm,algorithmic, multicol}

\usepackage{subfig}

\ifCLASSINFOpdf
  \usepackage[pdftex]{graphicx}
  \graphicspath{{./}}
  \DeclareGraphicsExtensions{.pdf,.jpeg,.png,.xps}
\else
\fi

\begin{document}

\title{VNF Placement with Replication for Load Balancing in NFV Networks}

\author{\IEEEauthorblockN{Francisco Carpio, Samia Dhahri and Admela Jukan}
	\IEEEauthorblockA{Technische Universit{\"a}t Braunschweig, Germany}
	\IEEEauthorblockA{Email:\{f.carpio, s.dhahri, a.jukan\}@tu-bs.de}
}

\maketitle

\begin{abstract}
Network Function Virtualization (NFV) is a new paradigm, enabling service innovation through virtualization of traditional network functions located flexibly in the network in form of Virtual Network Functions (VNFs). Since VNFs can only be placed onto servers located in networked data centers, which is the NFV's salient feature, the traffic directed to these data center areas has significant impact on network load balancing. Network load balancing can be even more critical for an ordered sequence of VNFs, also known as Service Function Chains (SFCs), a common cloud and network service approach today. To balance the network load, VNF's can be placed in a smaller cluster of servers in the network thus minimizing the distance to the data center. The optimization of the placement of these clusters is a challenge as also other factors need to be considered, such as the resource utilization.  To address this issue, we study the problem of VNF placement with replications, and especially the potential of  VNFs replications to help load balance the network. We design and compare three optimization methods, including Linear Programing (LP) model, Genetic Algorithm (GA) and Random Fit Placement Algorithm (RFPA) for the allocation and replication of VNFs. Our results show that the optimum placement and replication can significantly improve load balancing, for which we also propose a GA heuristics applicable to larger networks.
\end{abstract}

\IEEEpeerreviewmaketitle

\section{Introduction}

Network Function Virtualization (NFV) is a new paradigm that virtualizes the traditional network functions and places them into generic hardware and clouds, as opposed to the designated hardware. The placement of the virtual network functions (VNFs) can happen either in remote data centers or by deploying single servers or clusters of servers.  Placing VNFs in remote data center can lower the cost of deployment, but is known to typically increasing the delay and create churns of network load, due to the fix and often remote location. Installing new services (or, mini data centers) inside the network can mitigate the distance-to-datacenter problem. At the same time, the deployment of new servers forming small data centers in regular nodes requires new investment costs, which requires a gradual upgrade of the network.

To address the issue of whether the VNF should be placed in data centers or newly installed servers inside the network, solutions have been proposed to either placing: 1) minimum number of VNFs (i.e. minimum new hardware required), increasing the forwarding cost at the expenses of network traffic churns, or 2) maximum number of VNFs, thus decreasing the forwarding cost, while being able to balance the load better. In the second case, especially, assuming that network operator can deploy as many VNFs as needed, the traffic demands can be redirected to the closest VNF, decreasing traffic forwarding cost and solving the distance-to-datacenter problem. This solution, however, requires high initial deployment (and energy) costs. Therefore, there is a trade-off between the number of VNFs and forwarding costs that should be found in order to solve the so called, NFV resource allocation (NFV-RA) problem. While most of the current solutions minimize the number of VNFs under the resources' constraints, comparably less effort has been on addressing the network load balancing problem with VNF placement.

\begin{figure}[!t]
	\includegraphics[width=3.2in]{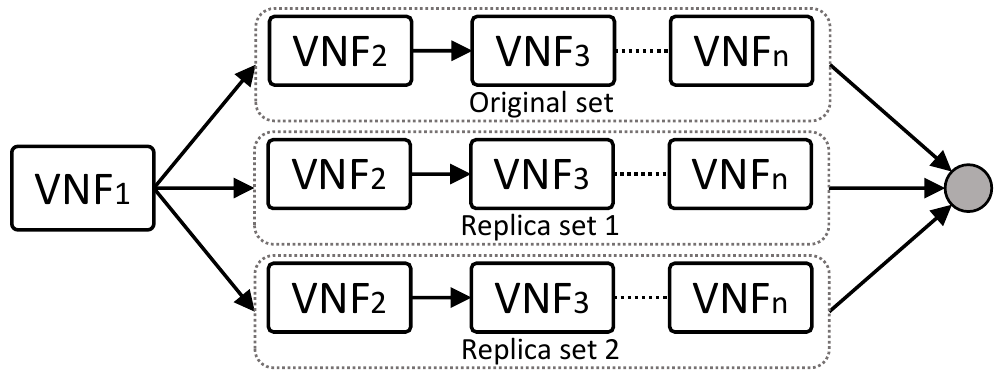}
	\caption{Service Function Chaining (SFC) with replications.}
	\label{sfc}
	\vspace{-0.4cm}
\end{figure}

In this paper, we study the problem of VNF placement with a novel concept of \emph{replications} to finding the solution for an optimum network load balancing. The VNF replicas in a service chain can be implemented as many time as needed in the network, such that optimum network traffic load balancing can be achieved when running a service. Fig. \ref{sfc} illustrates the idea, whereby we assume a service chain composed by one non-replicable VNF chained to one set and one or more sets of ordered sequence of VNFs, depending on the number of replicas, towards the service end-points. The non-replicable VNF is allocated in a dedicated data center which generates service requests, while the rest of the functions are allocated on small servers, maintaining the sequence order. To find the optimum placement of servers required for the deployment of VNFs, we formulate the problem as a Mixed Integer Programing (MIP) model. The model is compared with random allocation to demonstrate the necessity to optimize the placement for VNFs. For scalable optimizations that grow linearly with the size of the network, a Genetic Algorithm (GA) is proposed to find close-to-optimal solutions.

The rest of the paper is organized as follows. Section II presents related work. Section III describes the reference architecture. In Section IV and V, the related optimization model and the heuristics approaches are described, respectively. Section VI analyzes the performance, and Section VII concludes the paper and discusses future research.

\section{Related work}

Early work in \cite{Moens2015} studies the optimal VNFs placement in hybrid scenarios, where some network functions are provided by dedicated physical hardware and some are virtualized, depending on demand. They propose an ILP model model with the objective to minimize the number of physical nodes used, which limits the network size that can be studied due to complexity of the ILP model. In \cite{Mehraghdam2014}, a context-free language is proposed for the specification of VNFs and a Mixed Integer Quadratically Constrained Program (MIQCP) for the chaining and placement of VNFs in the network. The paper finds that the VNF placement depends on the objective, such as latency, number of allocated nodes, and link utilizations. In mobile core networks, \cite{Basta2014} discuss the virtualization of mobile gateways, i.e.,  Serving Gateways (S-GWs) and Packet Data Network Gateways (P-GWs) hosted in data centers. They analyze the optimum placements by taking into consideration the delay and network load. In \cite{Bagaa2014} also propose the instantiation and placement of PDN-GWs in form of VNFs.

Due to the inherent complexity of optimizations, heuristic or meta-heuristic algorithms have been proposed to finding near optimal solutions. Paper \cite{Cohen2015} minimizes the OPEX in the VNF placement problem separating into two NP-hard sub-problems, and proposing heuristic algorithms. Similarly, paper \cite{Xia2015} proposes heuristics to reducing computational complexity considering the resource demand in data centers. In \cite{Bari2016} two solutions are presented to the VNF-orchestration problem, an ILP model computing the optimal solution using CPLEX for small networks and a heuristic computing sub-optimal solutions for large networks.  Paper \cite{Rankothge2015} proposes a genetic algorithm for the VNF chain placement to satisfy the SLA and QoS objectives with dynamic traffic demands.
 
Similar to the previous work we use optimizations and heuristics to solve the VNF placement problem. Unlike previous work, we consider the replications, which is novel. Also, our approach is tailored to suiting the operational  mobile core networks, where the optimum placement of VNFs in data centers can be found based on maximizing the network load balancing, thus enabling a scalable growth of the mobile data traffic over years, critical to the emerging 5G networks. 

\begin{figure}[!t]
	\includegraphics[width=3.3in]{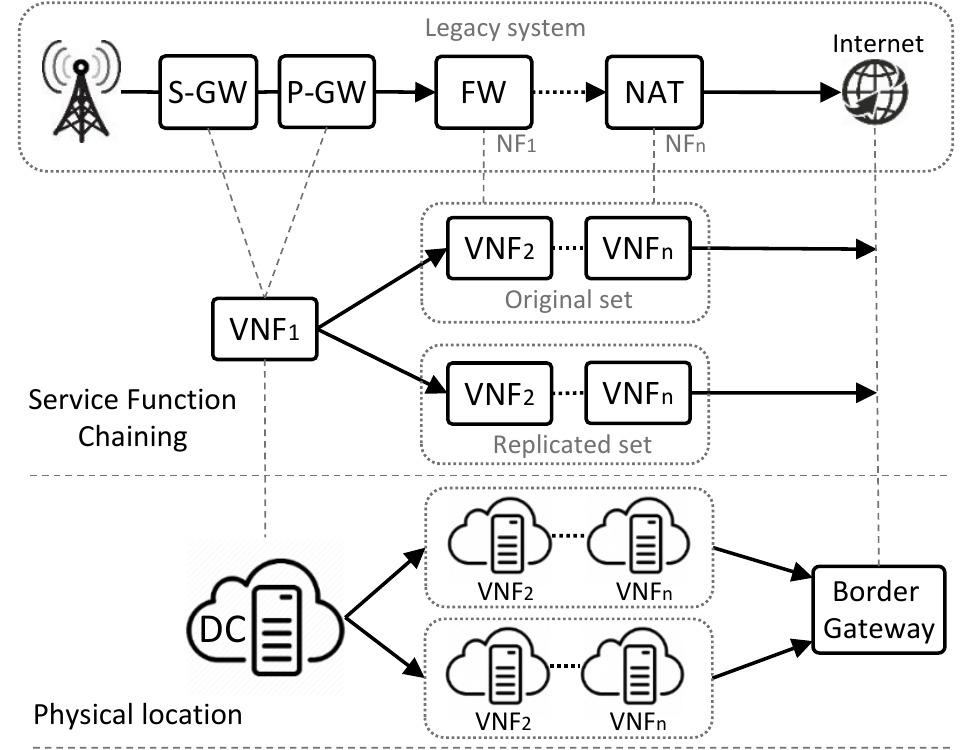}
	\caption{Use case on Mobile Core Networks}
	\label{use_case}
	\vspace{-0.3cm}
\end{figure} 

\section{Reference Architecture}

The NFV architecture is basically described by three components: Services, NFV Infrastructure (NFVI) and NFV Management and Orchestration (NFV-MANO). A \emph{Service} is the composition of VNFs that can be implemented in virtual machines running on operating systems or on the hardware directly. The hardware and software resources are provided by the NFVI that includes connectivity, computing, storage, etc. Finally, NFV-MANO is composed by the orchestrator, VNF managers and Virtualized Infrastructure Managers responsible for the management tasks applied to VNFs. In NFV-MANO, the orchestrator performs the resource allocation based on the conditions to perform the assignment of VNFs chains on the physical resources. Therefore, the NFV Resource Allocation problem \cite{Gil-herrera2016} is divided into three steps: 1) VNFs Chain Composition (VNFs-CC), 2) VNF Forwarding Graph Embedding (VNF-FGE) and 3) VNFs Scheduling (VNFs-SCH). The VNF-CC, also known as Service Function Chaining (SFC), studies the dynamic and strategic composition of VNF chains to be virtualized on physical network nodes. The VNF-CC problem is becoming more important in specially two areas where the IETF Network and Service Chaining Working Group is also contributing: \cite{draft-ietf-sfc-dc-use-cases-05} Data Centers and \cite{draft-ietf-sfc-use-case-mobility-06} Mobile Networks. The VNF-FGE challenge tries to find where to allocate the VNFs with regard to a specific objective (minimization of computation resources, minimization of power consumption, network load balancing, etc.). Finally, the VNFs-SCH schedules the execution of VNFs in order to minimize the total execution time of network services to improve the performance.

\subsection{Service Function Chaining (SFC) in Mobile Core Networks}

In the focus area of VNF-FGE, also known as VNF placement problem, proposed approaches focus on general cases, while less effort exist to show real use cases where an optimal VNF placement can be determinant. For this reason, we here propose to study a concrete scenario from legacy mobile systems, where we believe that the VNF placement in Service Function Chaining can bring most benefits. The proposed case study is shown in Fig. \ref{use_case}. The Serving Gateway (S-GW) and PDN Gateway (P-GW) are connected to e-NodeB and send the end-user traffic towards Internet. This traffic usually requires various additional services, currently deployed using traditionally embedded network functions, such as load balancers, TCP optimizers, firewalls and NATs. 

Considering the virtualization of the S-GW and P-GW on small data centers, as proposed in \cite{Basta2014}, we study the chaining and virtualization of the rest of functions on different physical locations in the mobile core network. To this end we define the problem as finding the optimum placement for these functions while at the same time load balancing the network. To answer this question, we propose to study where and how many NFV capable nodes are needed in the network for traffic load balancing. As illustrated at the bottom of Fig. \ref{use_case}, not only original functions, but also replicas of specific VNFs can be allocated in different network locations. The number of required replicas will be in relation with the network traffic demand. Therefore, by knowing how many replicas are necessary to maintain a good network load balancing, we know how many servers we need to build into the network. 

\subsection{Network Traffic Model}

\begin{figure}[!t]
	\includegraphics[width=3.3in]{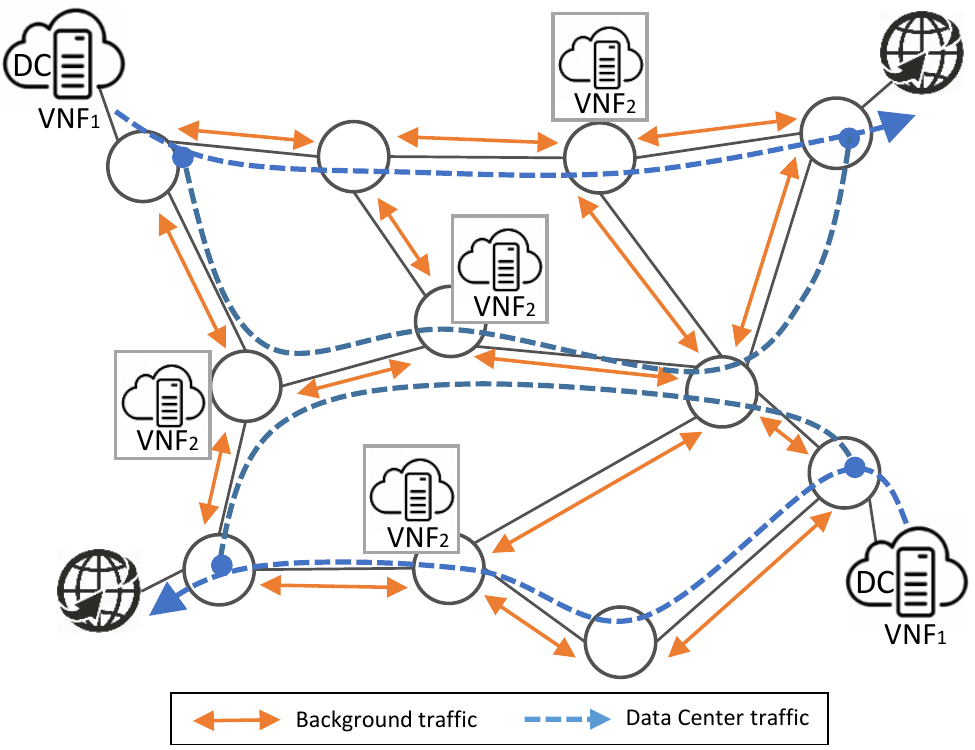}
	\caption{Network traffic considerations}
	\label{traffic}
	\vspace{-0.3cm}
\end{figure}

In this paper, we assume two kinds of network traffic, defined as the \emph{background traffic} and \emph{data center traffic}. This is illustrated in Fig. \ref{traffic}. The background traffic is the traffic generated from each core node to the rest of nodes and routed by a traditional network core protocol. The data centre traffic is generated by data centers, allocated in different parts of the network, towards the Internet gateways (Fig. \ref{use_case}). Since the latter category of traffic, usually TCP connections, is generated by end users, it has to traverse a set of network functions to match the required service before accessing to Internet.

In our approach, we assume that the background traffic can be generated randomly and forwarded following the rules of a specific Traffic Engineering (TE) model. The TE model written in form of ILP formulation (detailed in the next section) minimizes the link utilization of all links in the network using a linear cost functions approach. Because this traffic is not constrained by the VNF locations, the routing it is only constrained by traditional IP routing rules. Once the background traffic is load balanced, the output is used as an input parameter for the next model called Resource Allocation (RA). This model is used to allocate optimally VNFs in the network, while an optimal network load balancing is maintained. The optimum placement of VNFs and VNFs replicas can provide the optimum locations for the data centers, which will be responsible for the instantiation of VNFs.

\begin{table}[!t]
	\renewcommand{\arraystretch}{1.3}
	\caption{Notation}
	\label{parameters}
	\centering
	\footnotesize
	\begin{tabular}{c p{4.2cm}}
		\hline
		\textbf{Parameter} & \textbf{Meaning}\\
		\hline
		$\vv N = \{n_0, n_1, ..., n_{(N-1)}\}$ & set of all nodes\\
		$\vv L = \{l_0, l_1, ..., l_{(L-1)}\}$ & set of all links\\
		$\vv P = \{p_0, p_1, ..., p_{(P-1)}\}$ & set of all paths\\
		$\vv Y = \{y_0, y_1, ..., y_{(Y-1)}\}$ & set of linear cost functions\\
		$t_p^\ell \in \{0,1\}$ & 1 if path $p$ traverses link $\ell$\\
		$c_\ell$ & maximum capacity of link $\ell$ \\
		\multicolumn{2}{c}{\textbf{TE model}} \\
		\hline
		$\vv {\Lambda_{bg}} = \{\lambda_0, \lambda_1, ..., \lambda_{(\Lambda_{bg}-1)}\}$ & set of background traffic demands\\
		\multicolumn{2}{c}{\textbf{RA model}} \\
		\hline
		$\vv S = \{s_0, s_1, ..., s_{(S-1)}\}$ & set of service chains\\
		$\vv {V_s} = \{v_0, v_1, ..., v_{(V_s-1)}\}$ & set of VNFs in service chain $s$\\
		$\vv {\Lambda_s} = \{\lambda_0, \lambda_1, ..., \lambda_{(\Lambda_s-1)}\}$ & set of traffic demands for service chain $s$\\
		$r_{v} \in \{0,1\}$ & 1 if function $v$ can be replicated\\
		$r_{max} \in \{1,\infty\} $ & maximum number of allowed replicas per service chain\\
		\hline
		\textbf{Variable} & \textbf{Meaning}\\
		\hline
		$K_\ell$ & utilization cost of link $\ell$\\
		\multicolumn{2}{c}{\textbf{TE model}} \\
		\hline
		$R_p^\lambda \in \{0,1\}$ & 1 if traffic demand $\lambda$ is using path $p$\\
		\multicolumn{2}{c}{\textbf{RA model}} \\
		\hline
		$R_{p,s} \in \{0,1\}$ & 1 if service chain $s$ is using path $p$\\
		$R_{p,s}^\lambda \in \{0,1\}$ & 1 if traffic demand $\lambda$ from service chain $s$ is using path $p$\\
		$F_{s,v}^n \in \{0,1\}$ & 1 if VNF $v$ from service chain $s$ is allocated in node $n$\\
		\hline
		\vspace{-0.9cm}
	\end{tabular}
\end{table}

\section{Optimization Model}

\begin{figure*}[!t]
	\centering
	\subfloat[Nobel-us]{\includegraphics[width=3.5in]{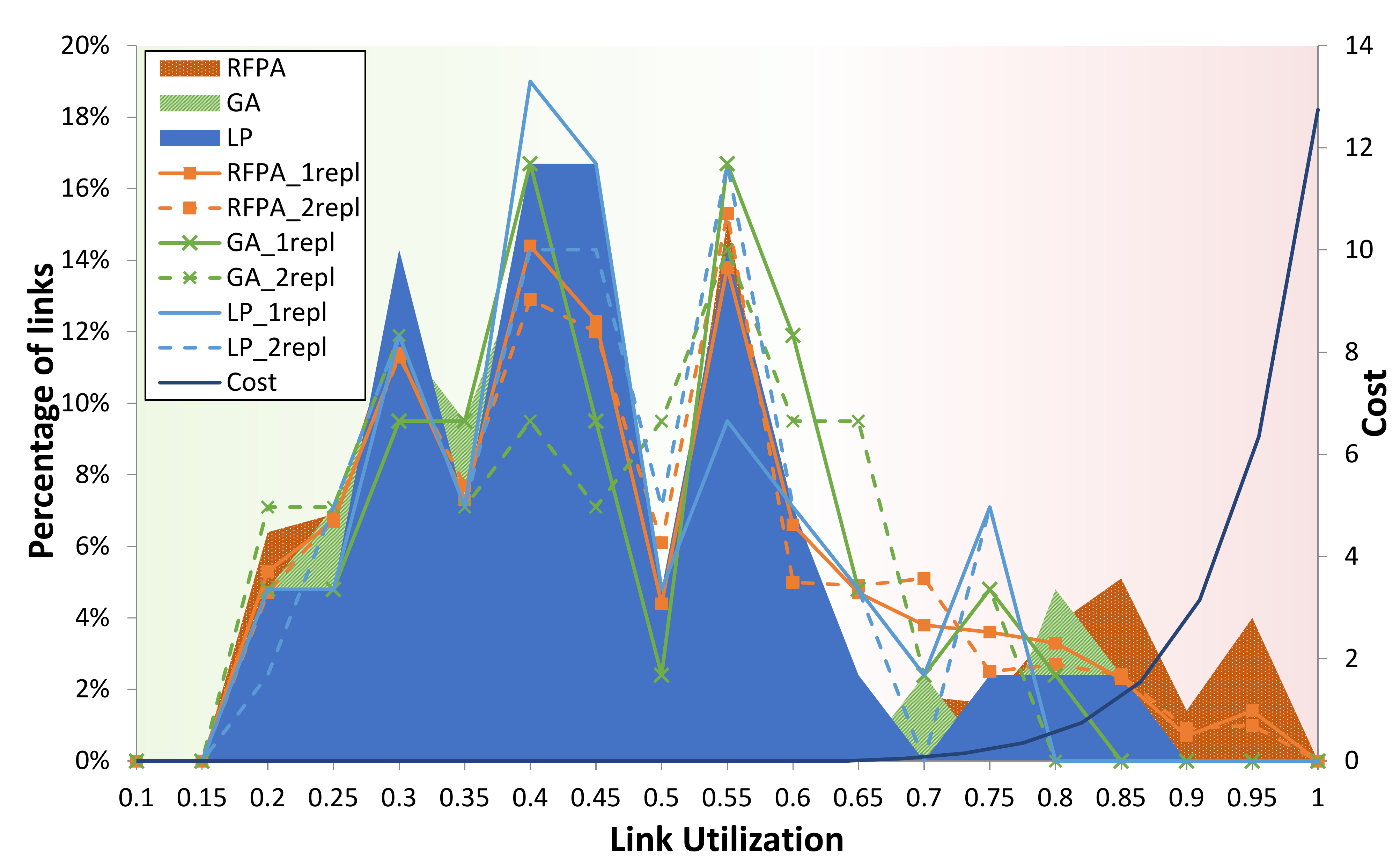}%
		\label{nsf}}
	\hfil
	\subfloat[Janos-us]{\includegraphics[width=3.5in]{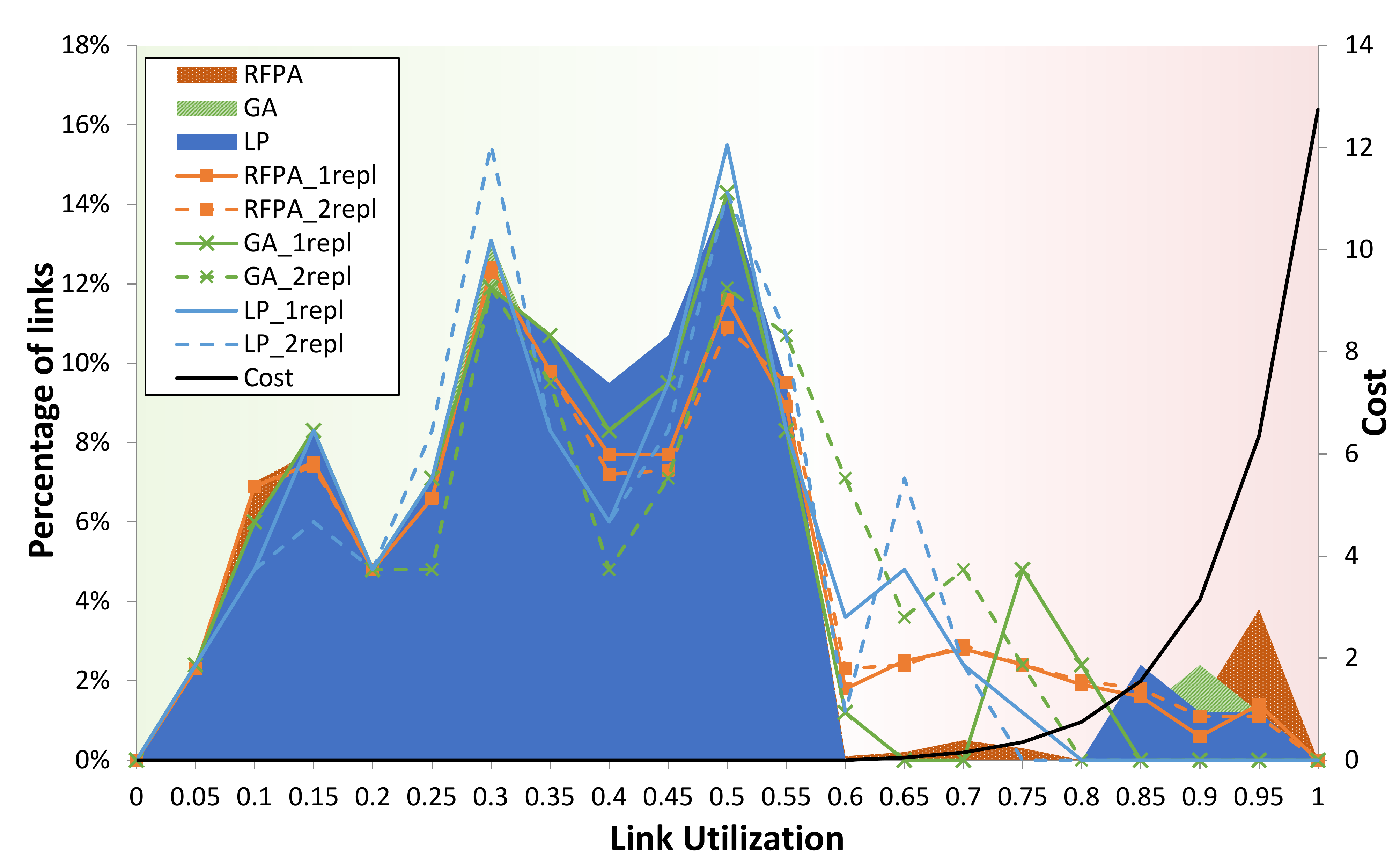}%
		\label{janos}}
	\vspace{-0.3cm}
	\caption{Link utilization results}
	\label{utilization}
\end{figure*}

This section formulates both the TE and RA models as optimization problems with the objective function that minimizes the cost for all links in the network, i.e., 

\begin{equation}
	Minimize: \sum_{\ell \in \vv L} K_\ell
\end{equation}

subject to a set of constraints, as described next. The notation of all parameters and variables is summarized in Table \ref{parameters}. The cost of every link is defined by the resulting value from all linear cost functions $y_i(U_{\ell}) = a \cdot U_{\ell} - b$, where $U_{\ell}$ is the link utilization specified by the term in the brackets. Then, $\forall \ell \in \vv L, \forall y_i \in \vv Y, \forall s \in \vv S, \forall p \in \vv P$:

\begin{equation}
	K_\ell \geq y_i \Bigg(\sum_{\lambda \in \vv \Lambda_{bg|s} }  \frac{\lambda \cdot (R_{p}^\lambda | R_{p,s}^\lambda)  \cdot t_p^\ell}{c_\ell}\Bigg)
\end{equation}

, where $R_{p}^\lambda$ or $R_{p,s}^\lambda$ are used in the TE or RA case, respectively. We assume zero cost for all links with utilization below 60\% and exponential increment cost between 60\% and 100\% (later shown in Fig. \ref{utilization}, \cite{caria}). The routing constraints for TE and RA models are, respectively:

\begin{equation}
\begin{split}
\forall \lambda \in \vv {\Lambda_{bg}}: \sum_{p \in \vv P} R_{p}^\lambda = 1 \\ \forall \lambda \in \vv {\Lambda_s}: \sum_{s \in \vv S} \sum_{p \in \vv P} R_{p,s}^\lambda = 1
\end{split}
\end{equation}

where, for both cases, the constraint assures that every traffic demand can only use one possible path.

The remaining constraints apply to the RA model only. The first constraint assures that a certain traffic demand $\lambda$ can only use a path $p$ only if the requested service chain is using the same path: 

\begin{equation}
\forall p \in \vv P, \forall s \in \vv S, \forall \lambda \in \vv {\Lambda_s}: R_{p,s}^\lambda \leq R_{p,s}
\end{equation}

For [$0, 1, 2, ..., r$] replicas, each service chain can use [$1,2,3...,(r+1)$] possible paths to forward traffic, i.e., 

\begin{equation}
	\forall s \in \vv S: 1 \leq \sum_{p \in \vv P} R_{p,s} \leq r_{max}
\end{equation}

Therefore, with increasing number of replicas in the network, we also increase the number of possible paths that a service chain can select to load balance the traffic. The next constraint allocates all VNFs from a specific service chain $s$ in the selected path:
\begin{equation}
\forall p \in \vv P, \forall s \in \vv S, \forall v \in \vv {V_s}: R_{p,s} \leq \sum_{n \in p} F_{s, v}^n
\end{equation}

The next constraint assures that two selected paths $p_1$ and $p_2$ from the service chain $s$ are not choosing the same location to place a certain intermediate function $F_{s, v}^n$ in any shared node $n$. For $\forall s \in \vv S, \forall p_1 \in \vv P, \forall p_2 \in \vv P, \forall v \in \vv {V_s}, \forall n \in p_1, p_2$:
\begin{equation}
R_{p_1,s} + R_{p_2,s} + 2 F_{s, v}^n \cdot r_v \leq 3
\end{equation}

Because the sequence order of VNFs in the service chain has to be maintained, in a selected path $p$, the function $v$ can not be allocated in the node $n$, if the previous function $v-1$ is not already allocated in any of the previous nodes of the same path. So, $\forall p \in \vv P, \forall s \in \vv S, \forall v \in \vv {V_s}, \forall n \in p$:

\begin{equation}
	\Bigg( \sum_{m = 0}^{n-1} F_{s, v-1}^m \Bigg) - F_{s, v}^n  \geq R_{p,s} - 1
\end{equation}

The remaining two constraints limit the maximum number of VNFs that can be allocated in the network. First, the maximum number of VNFs allocated in some specific node $n$ is constrained by:
\begin{equation}
	\forall n \in \vv N: \sum_{s \in \vv S} \sum_{v \in \vv {V_s}}  F_{s,v}^n \leq 1
\end{equation}

In other words, only one function can be allocated on node $n$. Second, if the function can be replicated $r_v$, then, the maximum number of replicas is constrained by $1+ r_{max}$. So, $\forall s \in \vv S, \forall v \in \vv {V_s}$:

\begin{equation}
\sum_n^N F_{s,v}^n \leq 1 + r_{max} \cdot r_v
\end{equation}

If the function is non-replicable (such as with S-GW/P-GW in the previous examples), the function can only be allocated once in the network per each service chain.  


\section{Heuristics}

Since the allocation of VNFs is known to be NP-hard \cite{Gil-herrera2016}, LP models are only feasible for small networks, while for large networks, heuristics are necessary to decrease the computation time. In this section we propose the use of a Genetic Algorithm (GA) for the placement and replication of VNFs just following the same procedure than the optimization model. In addition, we also propose to use a a Random Fit Placement Algorithm (RFPA). The purpose of random allocation is to know whether our approach has a considerable impact on the load balance, or alternatively simply building servers in the preferred node by the network operator is enough to load balance the network. 

\begin{algorithm}[!t]
	\caption{Genetic Algorithm}
	\begin{algorithmic}
		\renewcommand{\algorithmicrequire}{\textbf{Input:}} 
		\REQUIRE $\vv N, \vv L, \vv P, \vv Y, \vv S, \vv {V_s}, \vv {\Lambda_s}$
	\end{algorithmic}
	\begin{algorithmic}
		\STATE\textbf{Constraint}: 
		\begin{itemize}
			\item Background traffic - TE-GA
			\item Service Function Chain - RA-GA
			\item Number of replicas - RR-GA
		\end{itemize}
		\STATE\textbf{Initialization}: Population size, Individual definition
		\FOR{i=0 \TO num. of generations } 
		\STATE Evolve the population
		\FOR{j=0 \TO num. of individuals} 
		\STATE \textbf{Calculate fitness value:} Total Network Cost
		\ENDFOR
		\ENDFOR
		\STATE{\textbf{Output:}}
			\begin{itemize}
				\item Link utilizations, chosen paths - TE-GA
				\item Chosen nodes for allocation - RA-GA
				\item Chosen nodes for replication - RR-GA
			\end{itemize}
	\end{algorithmic}
	\label{ga}
\end{algorithm}

\begin{figure}
	\centering
	\includegraphics[width=3.2in]{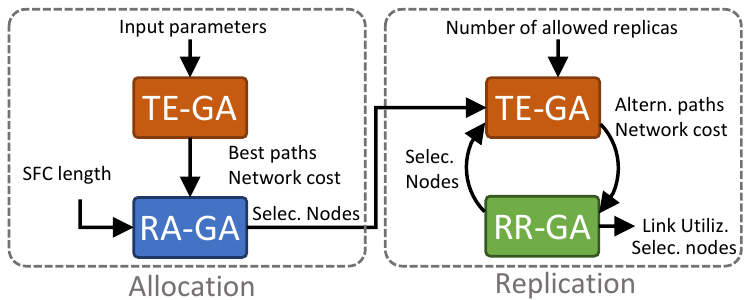}
	\caption{Genetic algorithm procedure}
	\label{ga}
	\vspace{-0.3cm}
\end{figure}

\subsection{Genetic Algorithm (Algorithm 1)}

The genetic algorithm is sub-divided into three interrelated genetic sub-algorithms: Traffic Engineering (TE-GA), Resource Allocation (RA-GA) and Resource Replication (RR-GA) algorithms. As shown in Fig. \ref{ga}, TE-GA algorithm selects a set of admissible paths based on the input parameters and calculates the network cost. The output is used as the input for the RA-GA algorithm which is responsible to allocate the original VNFs. The placement is carried out respecting the sequence order for the chosen admissible path, based on which placement produces a lower network cost after of routing the data center traffic. The selected nodes will be used as the input for replication, where with the maximum number of allowed replicas, the algorithm will try to find alternative paths. The alternatives paths are used in the RR-GA algorithm to allocate replicas based on the network cost, akin to RA-GA. Therefore, starting with one replica set, the network cost is checked and compared with the case without replication. If the cost decreases, then, the algorithm tries to allocate a second replica checking if the cost improves the previous case with one replica only. This procedure continues until the increment of the number of replicas can not anymore improve the cost. 

\subsection{Random-Fit Placement Algorithm (Algorithm 2)}

With this algorithm, the placement of VNFs and VNF replicas is carried out as random-fit, whereby all valid solutions according to the constraints defined in LP approach are considered and one of these solutions is randomly chosen. To find a valid TE and RA solution after the random placement of VNFs, the algorithm searches for the admissible paths that traverse the VNFs in the correct order and choose as many as the number of allowed replicas. Then, for each traffic demand, the algorithm selects one path randomly. The output will be the total network cost and the chosen nodes.

\begin{algorithm}[!t]
	\caption{Random Fit Placement Algorithm}
	\begin{algorithmic}
		\renewcommand{\algorithmicrequire}{\textbf{Input:}} 
		\REQUIRE $\vv N, \vv L, \vv P, \vv Y, \vv S, \vv {V_s}, \vv {\Lambda_s}$
		\renewcommand{\algorithmicrequire}{\textbf{Constraint:}} 
		\REQUIRE $\forall \vv L: C_l$
		\FOR{s = 0 \TO S } 
		\STATE \textbf{Do:} Place $\vv {V_s}$ randomly
		\ENDFOR
		\STATE \textbf{Do:} Find admissible paths
		\STATE \textbf{Do:} Choose random fit paths ($\leq r_{max}$)
		\FOR{$\lambda$ = 0 \TO $\Lambda_s$ } 
		\STATE Route $\lambda$ over one random path
		\ENDFOR
		\renewcommand{\algorithmicrequire}{\textbf{Output:}} 
		\REQUIRE Total Network Cost, chosen nodes
	\end{algorithmic}
	\label{rnd}
\end{algorithm}

\begin{table}[!t]
	\renewcommand{\arraystretch}{1.3}
	\caption{Parameters}
	\label{parameters_performance}
	\centering
	\scriptsize
	\begin{tabular}{l c c c c}
		\hline
		\textbf{Topology}  & \textbf{size (nodes-links)} & \textbf{conn.} & \textbf{DC-bw (Mbps)} & \textbf{BG-bw (Mbps)}\\
		\hline
		Nobel-us & 14-21 & 30 & 35 & (0, 160]\\
		Janos-us & 26-84 & 30 & 45 & (0, 50]\\
		Janos-us-ca & 39-122 & 25 & 50 & (0, 30]\\
		Germany & 50-88 & 25 & 35 & (0, 35]\\
		Ta2 & 65-108 & 20 & 45 & (0, 20]\\
		\hline
	\end{tabular}
	\vspace{-0.3cm}
\end{table}

\section{Performance evaluation} \label{performance}

In this section, we compare the LP model, implemented using the Gurobi Optimizer \cite{gurobi}, with the genetic algorithm and the random allocation approach. In Table \ref{parameters_performance}, all the analyzed topologies, chosen from SNDLib website \cite{sndlib}, are listed with the number of connections, data center bandwidth (DC-bw) and background bandwidth (BG-bw), respectively for each topology. In order to make the results comparable for all topologies, the number of data centers (i.e S/P-GW functions) is fixed to 2 and maximum link capacity to 2.5 Gbps. The background traffic is generated randomly with interval BG-bw, assuring that the cost generated by the TE model is always lower than 1. In other words, the background traffic does not create any capacity bottlenecks. The length of the service chain is composed by two end-points (e.g., S/P-GW datacenter and border gateway) and one intermediate VNF. We assume the location of border gateways as a random fixed parameter, while the location of the S/P-GW data centers and VNFs are the variables to optimizations. 

\begin{figure}
	\centering
	\includegraphics[width=3.1in]{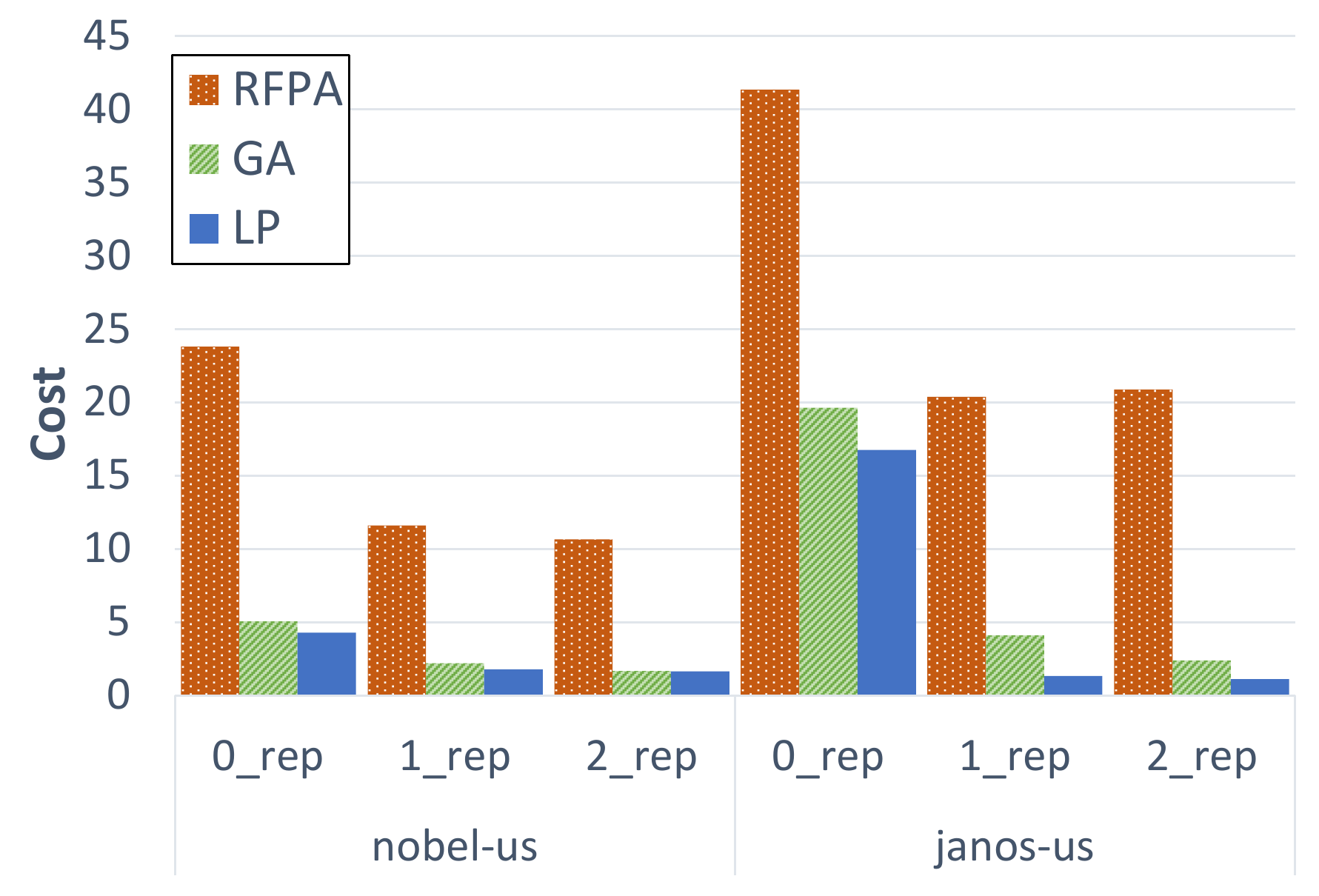}
	\vspace{-0.2cm}
	\caption{Network cost and number of replicas}
	\label{cost1}
	\vspace{-0.3cm}
\end{figure}

\subsection{Link utilization results}
In Fig. \ref{nsf} and \ref{janos}, we show the link utilization comparison between the LP model, Genetic algorithm and Random Fit Placement Algorithm, for Nobel-us and Janos-us topologies, respectively. With solid colors, each approach is represented without replicas, whereby the traffic of each S/P-GW data center has only one possible intermediate VNF to choose. We can appreciate how a random allocation of VNFs introduces a high number of overload links, while optimum and GA solution outperforms the results. With continuous and dotted lines we show the results for one and two replicas, respectively, for each approach. We can appreciate how introducing replicas without an optimum placement improves the case without replicas. Still, there is a considerable number of overload links. For optimum and near-optimum allocation, there are almost no links with utilization over 60\%.

\begin{figure}
	\centering
	\includegraphics[width=2.9in]{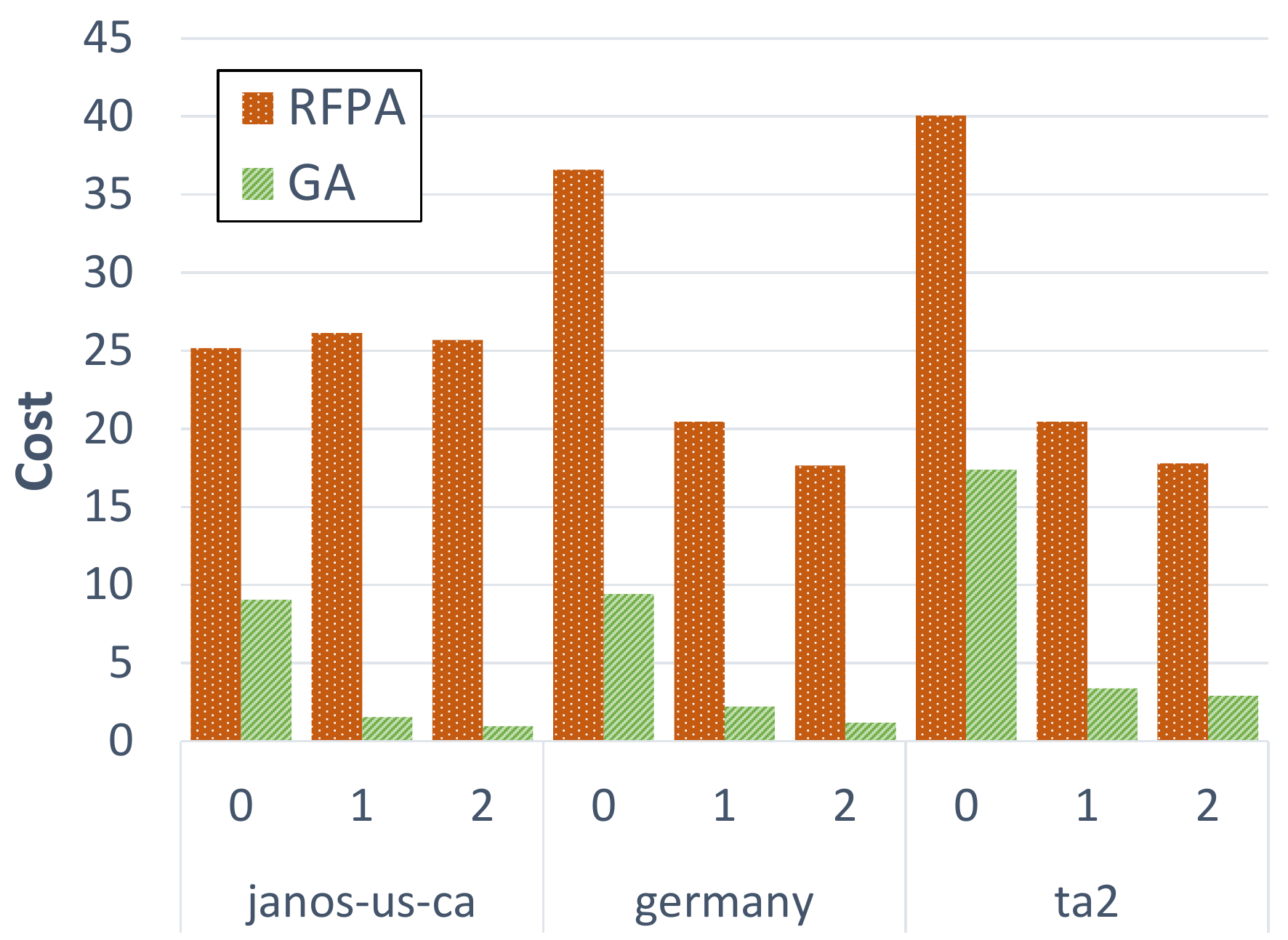}
	\vspace{-0.1cm}
	\caption{Network cost and number of replicas}
	\label{cost2}
	\vspace{-0.3cm}
\end{figure}

\subsection{Total network cost}

In Fig. \ref{cost1} presents the total network cost for two topologies, nobel-us and janos-us, and compares the three approaches. Here, we can see how the cost of the random allocation spikes in comparison with the genetic and the optimum solution, for both networks. The replication of VNFs also decreases the cost for all approaches, but performing specially better for the genetic and the optimum solutions. For large topologies, shown in Fig. \ref{cost2}, where LP model can not produce results in a reasonable period of time, the genetic algorithm is able to find solutions at much lower computational cost, specially in the case replications.

\subsection{Genetic algorithm benchmark}

\begin{figure}
	\centering
	\includegraphics[width=2.62in]{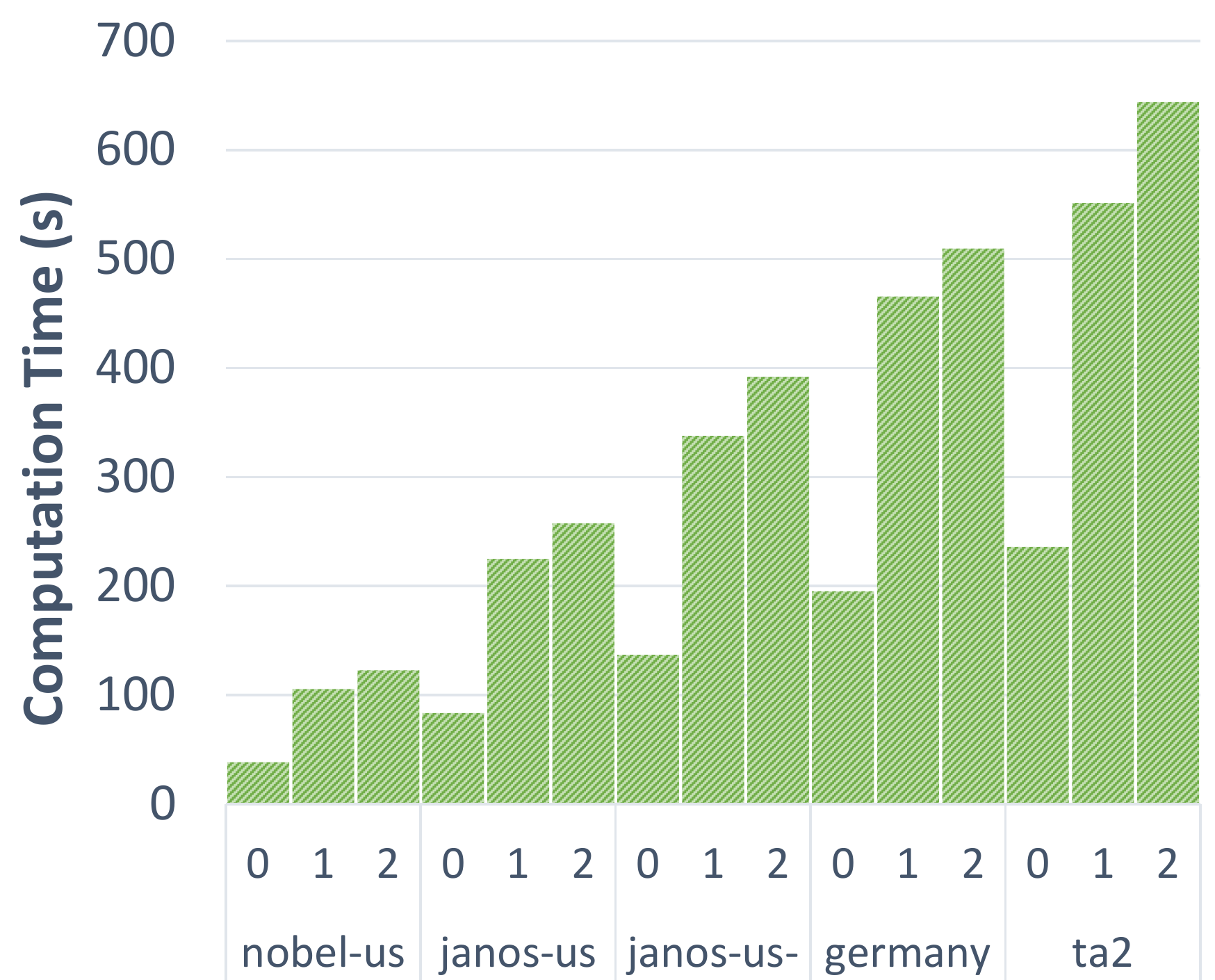}
	\vspace{-0.1cm}
	\caption{Computation time and number of replicas - GA}
	\label{time}
	\vspace{-0.3cm}
\end{figure}

The computation time of the genetic algorithm for different network sizes is represented in Fig. \ref{time}. As expected, for all topologies, the required computation time to find a valid solution is longer in replication cases due to the increment of admissible paths that every traffic demand can choose towards the destination. On the other hand, we appreciate how the GA complexity grows linearly with the topology size, and not exponentially as in the typical case of LP models. 

\section{Conclusion}

As VNFs can only be placed onto servers located in networked data centers, the traffic directed to these data center areas has significant impact on network load balancing and even more when this traffic has to traverse an ordered sequence of VNFs (service chain). To address this problem, VNF's can be placed in a smaller cluster of servers in the network solving the so-called  distance-to-data center problem. This motivated us to study the problem of VNF placement with replications in this paper, and especially how the \emph{replications} of VNFs can help to load balance the network. We designed and compared three optimization methods, including Linear Programming model, Genetic Algorithm and Random Fit Placement Algorithm. Our results show how the optimum VNF placement and replication in the network can significantly improve load balancing in comparison to simply building servers in the preferred nodes by the network operator. We also showed that heuristics can be effectively deployed to find near-optimal solutions with the the complexity that grows only linearly with the network size.

\bibliographystyle{IEEEtran}
\bibliography{mylib}

\end{document}